\input{aipcheck}

\documentclass[
    ,final            
  ]
  {aipproc}

\layoutstyle{8x11single}
\pdfoutput=1

\begin{document}

\title{Dynamics Underneath Symbols: A Case Study in Autonomous Agents}

\classification{89}
\keywords      {embodiment, T-maze, clock}

\author{Kohei Nakajima}{
  address={Artificial Intelligence Laboratory, Department of Informatics, University of Zurich}
}

\begin{abstract}
Our cognition is structuring the informational layer, consisting of perception, anticipation, and action, and it should also be sustained on a physical basis. In this paper, we aim to explore the relationship between the informational layer and the physical layer from a dynamical systems point of view. As an example, the fluctuation of choice is investigated by using a simulated agent. By setting a T-maze, the agent should choose one arm of the maze if a corresponding token is presented. We prepared two types of tokens, corresponding to the left and right arm of the maze. After training the network of the agent to successfully choose the corresponding arm, we presented two tokens simultaneously to the agent and observed its behavior. As a result, we found several behaviors, which are difficult to speculate on from a case in which only a single token is presented to the agent. Detailed analyses and the implications of the model are discussed.
\end{abstract}

\maketitle


\section{Introduction}
The aim of this paper is to explore how the informational process is sustained in the temporal manner in cognition. Our daily cognition is considered to be structuring the informational layer, consisting of perception, anticipation, and action. For example, when a man sees some specific token or symbol drawn or written on the wall, he might immediately start running. This phenomena is difficult to explain only from a mechanical point of view. Facing these phenomena, we usually tend to consider that there exists a meaning for the symbol, and because he understood the meaning, he immediately started to run. Recently, in embodied robotics \cite{RolfUnderstanding}, it is suggested, however, that the thing we call meaning is not statically determined by internal representations but constantly sustained and self-organized by dynamical interactions between our brain, body, and environment \cite{RolfScience,RolfHow,Andy,noe}. Concerning this point, a huge number of model studies have been done using embodied autonomous robots or artificial agents \cite{RolfUnderstanding,RolfScience,RolfHow,Oregan}.

In this context, Morimoto and Ikegami \cite{Morimoto} constructed an agent that could discriminate between the different shapes of tokens. In this model, the discrimination was not given explicitly in the top down manner, but was dynamically accompanied by the exploring behavior of the agent in the environment. They showed that the agent was switching between three different phases autonomously in discriminations. One was the wandering phase, which wandered the environment to search for tokens. The second was an examining phase, which examined the encountered token, and the third was the filling phase, which decided the shape of tokens. It suggested that the constant interaction between the embodied agent and the environment generated what we call the meaning of tokens. Since we cannot find a static internal representation as a whole, especially in this case, we could say that what we call meaning might be temporally stretched, decomposed, and distributed in each dynamics.

\begin{figure}[htbp]
	\centering
	\includegraphics[width=5.2in, bb=21 320 562 642]{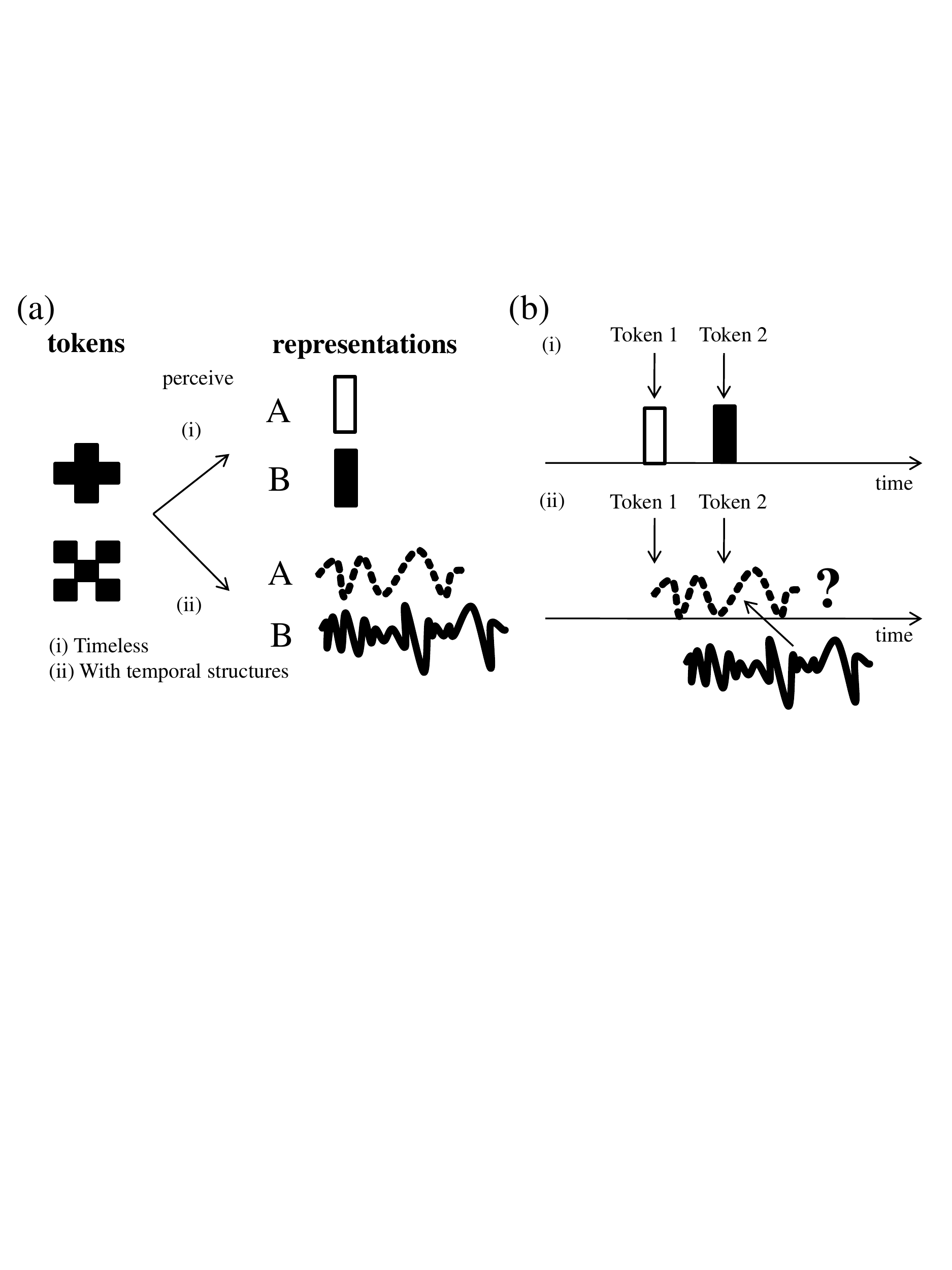}
	\caption{(a) When the agent perceives the tokens in the environment as representations, they are represented as timeless (i). On the other hand, when the agent perceives the tokens as dynamics originate from the interaction between brain, body, and environment, they are stretched in a temporal dimension as an internal dynamics (ii). (b) Assume that the agent perceives two tokens in order, if it perceives them as representations, since they are timeless, they can coexist in the physical time (i). But if it perceives them as dynamics, since they posses certain durations of time to process, we cannot guarantee whether they can coexist in the physical time or not (ii).}
	\label{fig0}
\end{figure}

What we focus on in this paper is concerned with the temporally stretched informational process in the interacting simulated autonomous agent. We constructed the simple model agent, which behaves as if it understands the different kinds of symbols and reveal their latent temporal structure. Next, we examined the natural consequences we might face when we pressed this line of thought. If we assume that the meaning is accompanied by the internal representations, because the representation is itself timeless (Fig. \ref{fig0} (a)), we don't need to consider an intervention of each other even if those symbols were presented in relatively short succession (Fig. \ref{fig0} (b)). But if the meaning was temporally stretched, how can those meanings sustain each other or be destroyed in the agent (Fig. \ref{fig0} (b)). 

To see this, we adopted a simple T-maze setting \cite{Tom,Tani}. Maze studies are often used to study spatial learning and memory in rodents. They have helped uncover general principles about learning that can be applied to many species. 
A T-maze is a type of maze which is shaped like a ``T''. The test animal starts at the base of the T. Depending on what we want to observe from the experiment, setting some conditions, the animal should walk forward and choose the left or right arm of the maze. Now, let us consider that we let the embodied agent learn to go left when it perceives some token, and go right with the other token. What happens when the two tokens are presented simultaneously? For example, when the token that means go left was presented, the temporal process for going left would start to drive the agent. But before it adjusts the process completely, if it perceives the token to go right, the token might work as a perturbation to the first token, or it might even start to go right.
In this setting, what we prepared is a meaning for each token and how to learn them, and so it is not trivial which direction the agent chooses in this situation.
The choice that the agent makes in this situation is called, in this paper, an autonomous choice of a model agent.
The term ``autonomous'' is used in various contexts, so it would be best to clarify and limit the use of it. 
Since we, the designers, are modeling an agent, its functions are somehow already determined as a specific dynamical system. Thus, in the strong sense, our agent is not autonomous at all \cite{Kohei2,Kohei4,niizato}. On the other hand, although its functions are determined, the designers of the agents should adopt a certain optimization method to obtain them. If we deal with the functionality of the agent, which was not explicitly intended to be optimized, then it can be said that it was not determined by the designers, so the functionality of the agent is autonomous from the designers \cite{Kohei3}.
Here, we restrict the use of the term as the latter case.
By using these settings we aim to explore how the informational process of a sensorimotor coupling system is sustained in the temporal manner.

This paper is organized as follows. In the next section, we explain the detailed model description of our simulation experiment, and then we explain our results and analyze the temporal structure, which our agent accompanies, and observe the autonomous choice in our agent. Finally, we will discuss some of the implications of this study.

\section{Model Description}
\subsection{T-maze, Network Architecture, and Tasks}
\begin{figure}[htbp]
	\centering
	\includegraphics[width=5.5in, bb=16 327 573 622]{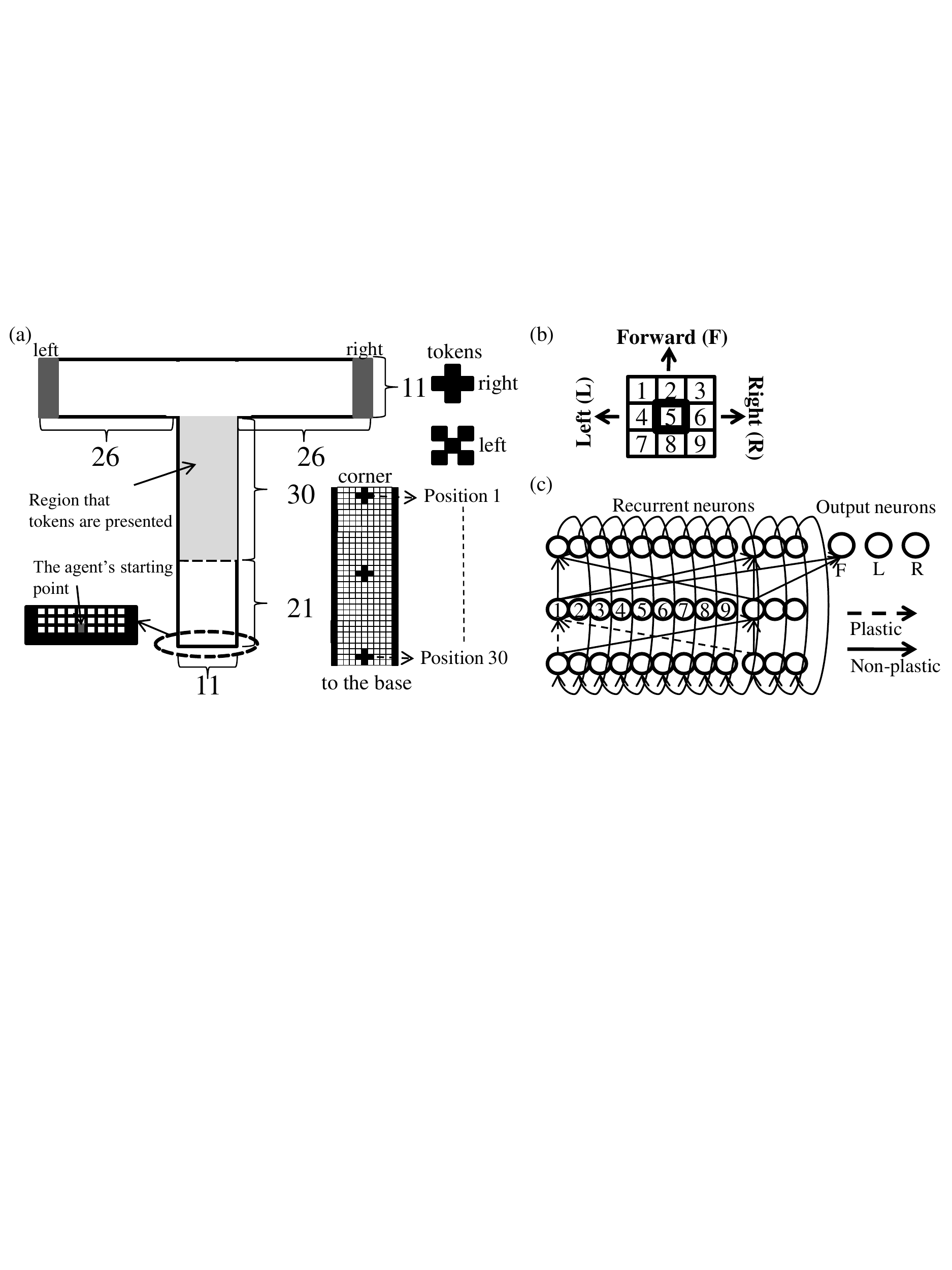}
	\caption{(a) The settings of T-maze, tokens, and the localization of tokens adopted in this paper. (b) Outer image of an agent. (c) The network architecture of the agent. From the input layer to the middle layer, and from the middle layer to the output layer, are totally connected (not all the connections are presented in the figure). Comparing the values of three output neurons, the largest one is adopted for the motor output in each timestep. The plasticity is introduced only for the weights of connections from the input layer to the 9 specific neurons in the middle layer. See text for details.}
	\label{fig1}
\end{figure}

Fig. \ref{fig1} (a) shows the maze, tokens, and the region where those tokens are localized.
The T-maze is expressed as a two-dimensional discrete lattice space, and each point is in one of two states, empty (0) or occupied (1). For the types of token, we adopted two in this paper: token R and token L, which mean ``go right'' or ``go left'', respectively. We use occupied cells to express both the wall of the maze and the tokens. Inside the T-maze, there is a specific region where tokens are presented. The region stretches for 30 cell lines downward from the crotch of the maze. Tokens are centered in the arm, and can take various arrangements in a vertical direction. There are 30 possible patterns of arrangement for one token, and we index the position from up to down, as from Position 1 to Position 30, for descriptive purposes.

An agent is situated in a point inside the T-maze and starts exploring the field from the base. It receives sensory inputs from the point it stays and eight neighboring points. Therefore, nine bits of information, in total, can be used to decide the next movement. On every discrete time step the agent changes the position to a neighboring point, according to the motor output. The motor output can be any of the three directions, namely forward, left, or right. Note that our agent does not have headings, which means that the direction of the agent's movement corresponds to that of the field (Fig. \ref {fig1} (b)). And moreover, it cannot go across the wall (if the agent was situated at the right hand side of the wall and the agent's next movement was to go left, it stays at the same point).

The agent is equipped with a recurrent neural network with plasticity to decide the next movement from the current state. Fig. \ref{fig1} (c) depicts the internal network structure. It consists of three layers, namely the input layer (12 neurons), the middle layer (12 neurons), and the output layer (15 neurons). In the middle layer, nine specific neurons receive the rare sensori-input directly. The activation of neurons is updated internally as follows:
\begin{equation}
	y_{i} (t) = g( \Sigma_{j} w_{i j} y_{j}' (t-1) + b_{i}),
\end{equation}
\begin{equation}
	g(x) = (1+ \exp (-\beta x))^{-1},
\end{equation}
where $y_{i}' (t)$ is the actual activation of the $i$ th neuron at the time $t$, $y_{i}(t)$ is the value generated by the internal dynamics, $w_{i j}$ is the weight of connection from the $j$ th neuron to the $i$ th neuron, and $b_{i}$ is the bias of the $i$ th neuron. The summation is taken of all neurons, which have a connection with the $i$ th neuron. $g(x)$ is the sigmoid function. As a result, the activations take the value from (0,1). $\beta$ is the nonlinearity coefficient and is set to 1.0 in this paper.
Actual activations of neurons are modified as follows:
\begin{equation}
	y_{i}'(t)= (1-\mu)s_{i} (t) + \mu y_{i}(t),
\end{equation}
where $s_{i}$ is the raw sensory input and takes one of two values, namely 0 or 1. This procedure is adopted only for the neurons that receive the rare sensori-input, namely the nine specific neurons in the middle layer, and otherwise $y'(t)=y(t)$. The meaning of Eq. 3 is that the actual activations of input neurons are modified by the value that is generated from the internal dynamics. In this paper, we used $\mu$=0.3. Therefore, the activation of an input neuron takes a value from (0,0.3) or (0.7,1.0), depending on the state of the corresponding position in the field.
In addition, we introduced a plasticity of weights. All the weights of connections, from the input layer into the nine specific neurons that receive the sensori-input directly, change during the interaction with the environment. These weights are updated as follows:
\begin{equation}
	\Delta w_{i j} (t) = \eta_{i} (s_{i} (t) - y_{i} (t) ) y_{j}'(t-1),
\end{equation}
where $\eta_{i}$ is the learning rate. By introducing the plasticity in this way, the value $y_{i}$ corresponds to the input neurons, which is generated from the internal dynamics, and has the tendency to approach $s_{i}$.

\subsection{Genetic Algorithm}
To train the network, we adopted three tasks. For task 1, token L is presented, and the agent should perceive the token and get to the end of the left arm. For task 2, token R is presented, and the agent should perceive the tokens and get to the end of the right arm. And for task 3, both token L and token R are presented, and this agent should perceive both tokens and get to the end of either the left or right arm. So in the case of task 3, we can determine the autonomous choice of the agent. A similar setting can be found in \cite{Ogai}.

For each task, if the agent successfully perceives the token, it gets 50 points, and if the agent eventually gets to the corresponding side of the arm, it gets 100 points (in the case of task 3, if it gets to either side of the arm, it gets 100 points), and every time it crushes to the wall, it gets -5 points. For each task, we run 10 trials of simulations and sum up all the points as a score. The agent can proceed to task 3 only if the score of task 1 and task 2 exceeds 0. In each trial, the performance of every agent is evaluated in a different token arrangement randomly chosen from the 30 possible positions for each token. The movement is simulated for 500 timesteps, and if the agent gets to the left or right end earlier than 500 timesteps, the simulation stops at that timestep. 

By evaluating the agent's behavior according to the score, we determined the agent's weights, biases, and the learning rates of neural networks. The number of population is 1000. The best agents are carried over to the next generation without any modifications. Other agents are generated from the best agents by adding small random values from the range [-0.01:0.01] to the agent's weights, biases, and the learning rates; no crossover is performed. In the case of plastic connections, values are interpreted as the initial values of the weights. Weight values of plastic connections are reset to the initial values when the agent starts the new trial, and all the initial recurrent input is set to 0.5 each time.

\section{Results}
\begin{figure}[htbp]
\centerline{\includegraphics[width=4.5in, bb=83 312 450 631]{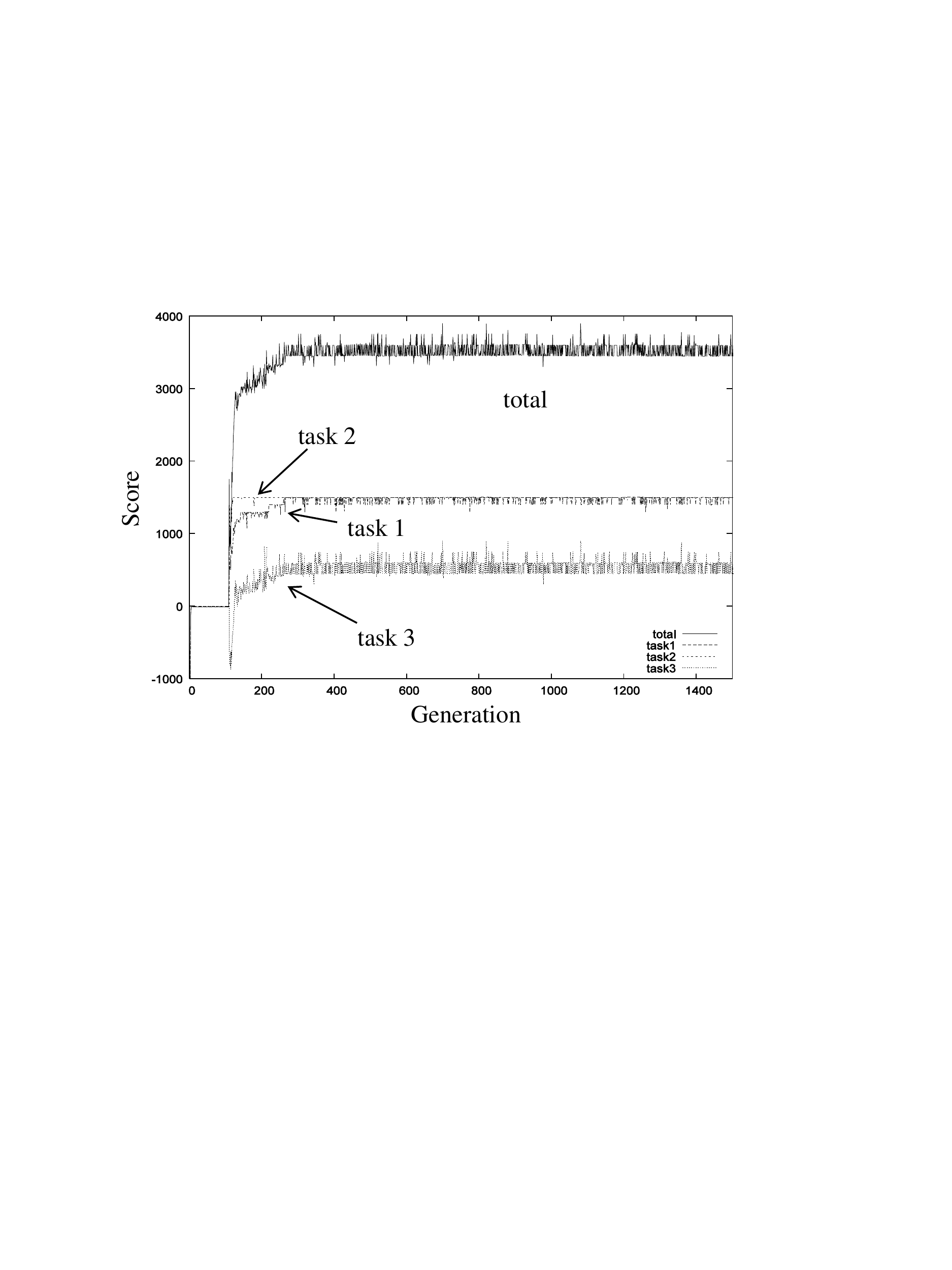}} 
\caption{Scores of the best individual from each generation. The score of task 1, task 2, task 3 and total score are presented.} \label{fig3}
\end{figure}

Fig. \ref{fig3} shows the increase of the score of the best individual from each generation. By using the best individuals from generation 100 (Agent 1) and generation 1500 (Agent 2), we can analyze their behaviors, the strength of the bundles of those behaviors, and discuss the autonomous choices they accompany.
 
\subsection{Observations}
\begin{figure}[htbp]
	\centering
	\includegraphics[width=5.5in, bb=19 204 575 615]{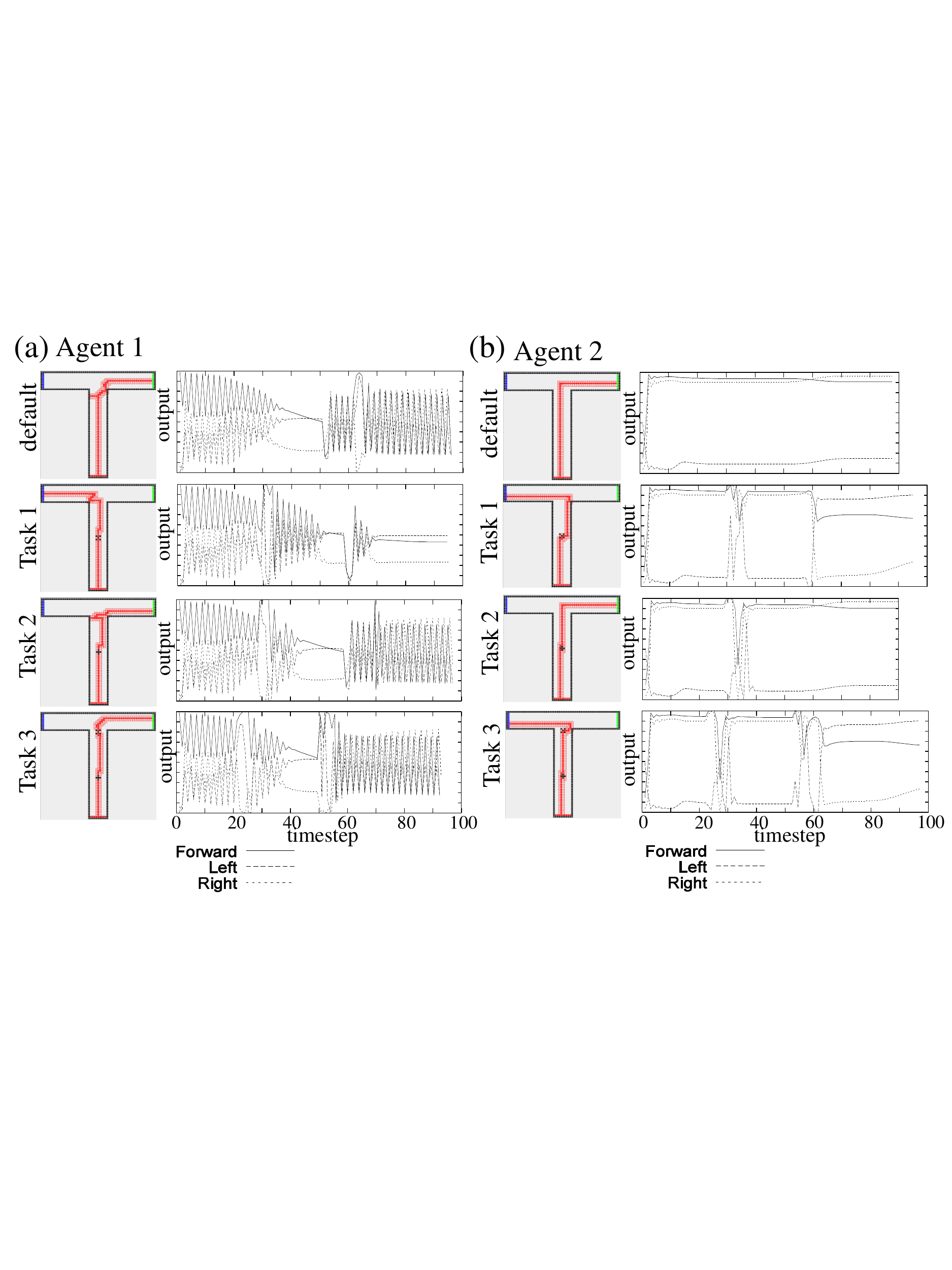}
	\caption{Typical behaviors of Agent 1 (a) and Agent 2 (b) for the default condition, task 1, task 2 and task 3. For each line, the left figure shows the agent's trajectory in the T-maze and the right diagram shows the dynamics of its outputs.}
	\label{fig4}
\end{figure}

Fig. \ref{fig4} (a) and (b) show trajectories in the T-maze and the dynamics for the outputs of Agent 1 and Agent 2, respectively.
They show results for default conditions (without tokens), task 1, task 2, and task 3.
By observing default conditions, we can observe the agents' natural side preferences in the T-maze.

Let us see the behaviors of Agent 1 (Fig. \ref{fig4} (a)). When the tokens are not presented, all the dynamics of outputs are periodic, maintaining the value of the forward output as the largest. And then the dynamics of outputs stop to show the periodic behavior, and in timestep 49 the left output exceeds the value of the forward output. In timestep 50, the agent crushes to the left wall, and in timestep 61 it crushes to the right wall, and at last the dynamics of the outputs start to show periodic behavior again, maintaining the value of the right output as the largest and arriving at the tip of right arm. In task 1, task 2, and task 3, we can see that the dynamics of the outputs of the default condition are dominant.
Moreover, we can see that the agent is using the wall effectively to achieve the task.

Next, let us see the behaviors of Agent 2 (Fig. \ref{fig4} (b)). In the default condition, the agent goes straight forward to the right arm without crushing the wall. And the dynamics of output are calmly stable, maintaining the value of the forward output as the largest till timestep 59, and after that the right output shows the largest. In task 2, when the right token was presented, we can see that the agent uses this property of the default conditions. On the contrary, if the left token was presented, it uses the right wall to change the direction. What is interesting is that it shows the same pattern of perception whenever the token is presented. We can confirm this from the dynamics of outputs.

Comparing Agent 1 and Agent 2, we can say that the behavior of Agent 1 is relatively up to its internal dynamics, so that the effects of the tokens are largely dependent on the timing and phases of the internal dynamics. It means that even though the token is the same, its effect on the agent is not always the same. On the contrary, Agent 2 has moderately stable behavior, so that whenever it perceives the token, it reflects as the same.

\subsection{The Bundle of Behaviors and the Autonomous Choice}
The behavior of a dynamical system is dependent on the initial states. Hence, in this paper, to compare the agents' dynamical behaviors, we have evolved the agent to distinguish the left and right tokens by daringly fixing all the initial recurrent inputs to 0.5 for each case. By doing this, we can compare the relevance of the learning instead of its initial conditions. In the real animal case, its behavior might be highly dependent on its history, and the realized trajectory should be one of many possible trajectories. In this subsection, examining the behavior of the agents by applying the range of noise to the initial recurrent inputs, we evaluate the sustainability of the learned trajectories. This is aimed at seeing the hidden structure, the bundle of behaviors that the agents contain. By varying the strength of noise, $\epsilon$, in incremental steps, we calculated the judgment probability for left, right, and showing no judgment in the various token arrangements in the default, task 1, task 2, and task 3 cases. For noise, we applied $0.5 \pm \epsilon$ to each initial recurrent input, where $\epsilon$ is a random real value whose range is tested incrementally for $[0.0, 0.1]$, $[0.0, 0.2]$, $[0.0, 0.3]$, $[0.0, 0.4]$, and $[0.0, 0.5]$.
All the results of judgment probabilities are obtained by averaging the behavior of the agent for 100 runs in each setting.

\begin{figure}[htbp]
	\centering
	\includegraphics[width=5.2in, bb=116 500 437 677]{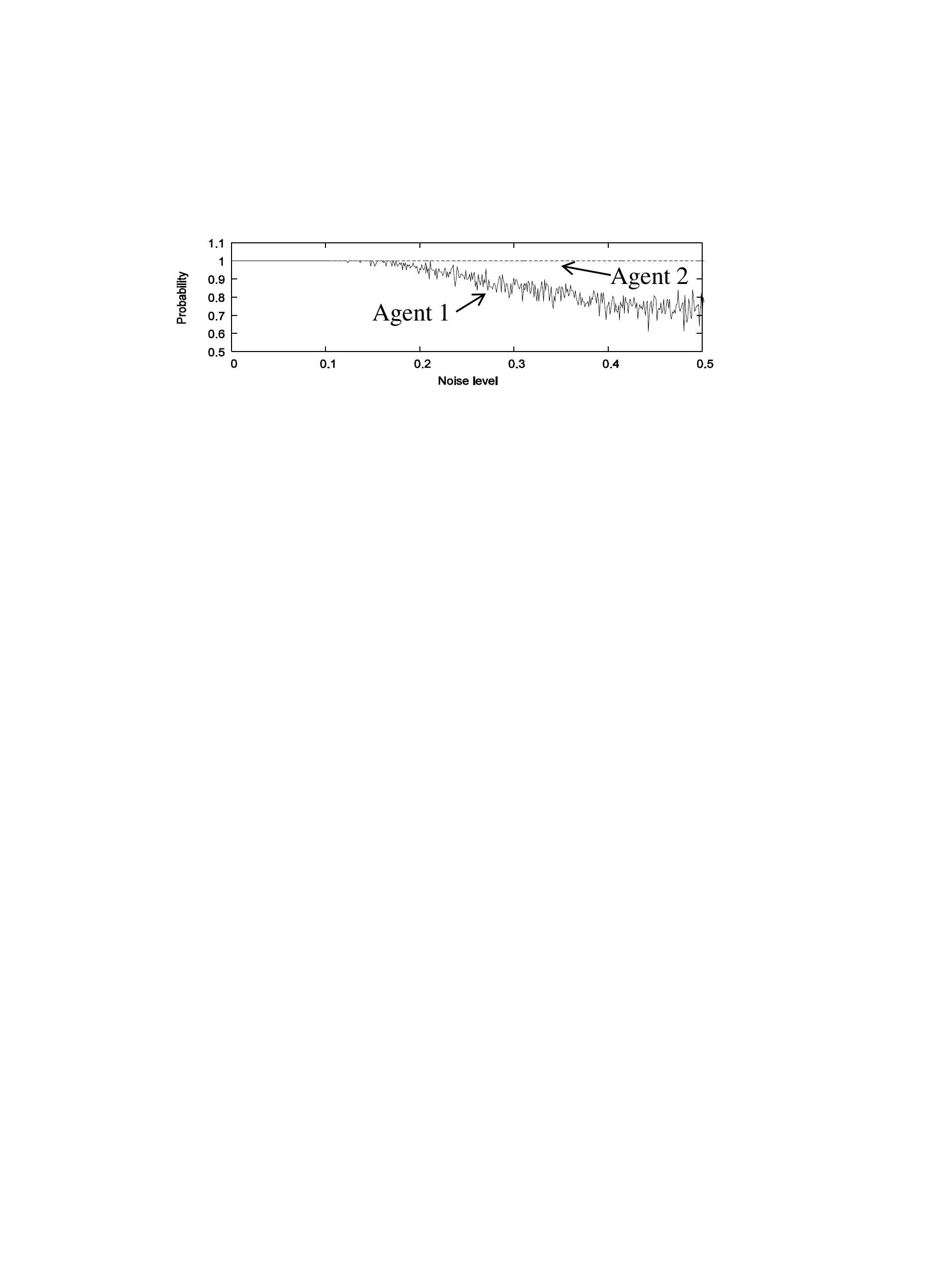}
	\caption{Probability to get to the end of right arm in the default conditions for Agent 1 and Agent 2 as a function of noise levels. See texts for details.}
	\label{fig5}
\end{figure}

In Fig. \ref{fig5}, we observed the relevance of noise to the side preference in the default condition.
In Agent 1, by increasing the noise level, the rate, which showed 1.0 to go to the right arm, was decreased little by little, and when the noise level was maximum value, it showed 0.6 (0.4 of the trial will go to the left arm). On the other hand, in Agent 2, the side preference would not change even if the noise level was at maximum value.

\begin{figure}[htbp]
	\centering
	\includegraphics[width=5.5in, bb=19 257 556 628]{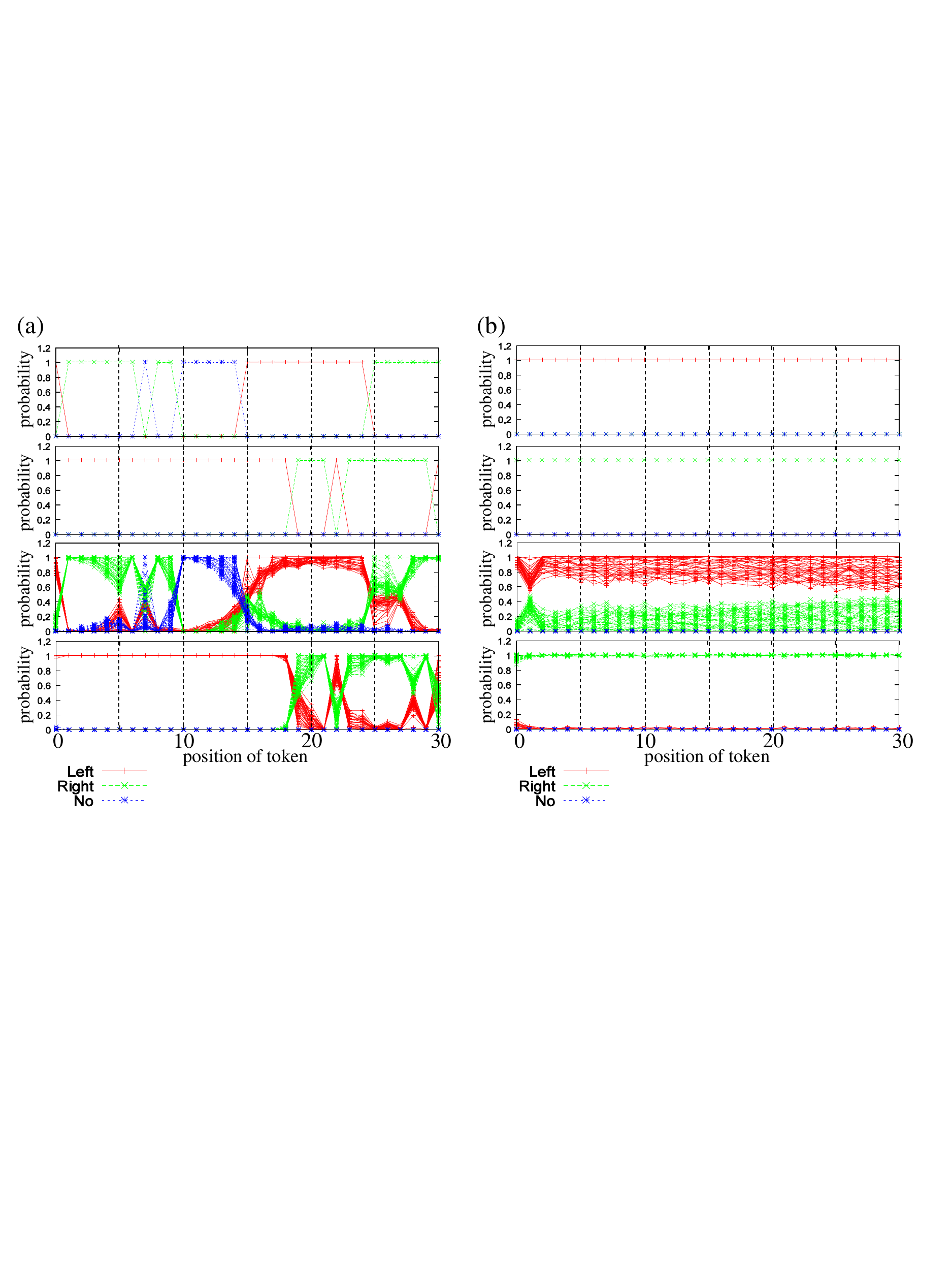}
	\caption{Probability to get to the end of the left arm, the right arm, or not to get to either of the arms, as a function of the position of the tokens. (a) is the result of agent 1, and (b) is that of agent 2. The upper two diagrams are the results without noise, and the lower two diagrams are the results with noise, for task 1 and task 2, respectively. Noise levels are determined for five incremental steps, and the results are overlaid for all the noise levels.}
	\label{fig6}
\end{figure}

Next, we examined the relevance of noise to task 1 and task 2 (Fig. \ref{fig6}). Let us see the cases in Agent 1 (Fig. \ref{fig6} (a)). In task 1, when there was no noise, if the token was in position 0, 15-24, then they showed a correct judgment. If the token was in position 1-6, 8-9, or 25-30, they showed an opposite judgment. And if the token was in position 7 or 10-14, they showed neither judgment. In task 2, when there was no noise, if the token was in position 19-21 or 23-29, then they showed a correct judgment. If the token was in position 0-18, 22, or 30, they showed an opposite judgment. And there was no case that showed neither judgment. On the contrary, when there was a noise, it was influenced when the tokens were in specific arrangements. Namely, in task 1, when the token was in position 5, 7, 9, 15, 16, 25, 26, or 27, and in task 2, in position 19, 20, 22, 27, 28, or 30. When the tokens were in different arrangements, it did not influence the original judgments. This result implies that Agent 1 contains temporal structures. Because it means that the timing for perceiving the tokens is crucial, and the influence from the noise is not equal among all the token arrangements and shows specificity for each token arrangement. This result is difficult to predict only from the results shown in Fig. \ref{fig5}, since it does not contain the timing for perceiving the token.

In Agent 2 (Fig. \ref{fig6} (b)), when there was no noise, it shows completely correct judgments for both task 1 and task 2. When we increase a noise level, the judgment probability gradually decreases for task 1. And in task 2, the judgment probability is not as influenced by the noise. These results show that its judgment is not dependent on the position of the tokens, hence it is possible to say that does not contain a temporal structure. 

\begin{figure}[htbp]
	\centering
	\includegraphics[width=6.0in, bb=21 323 570 585]{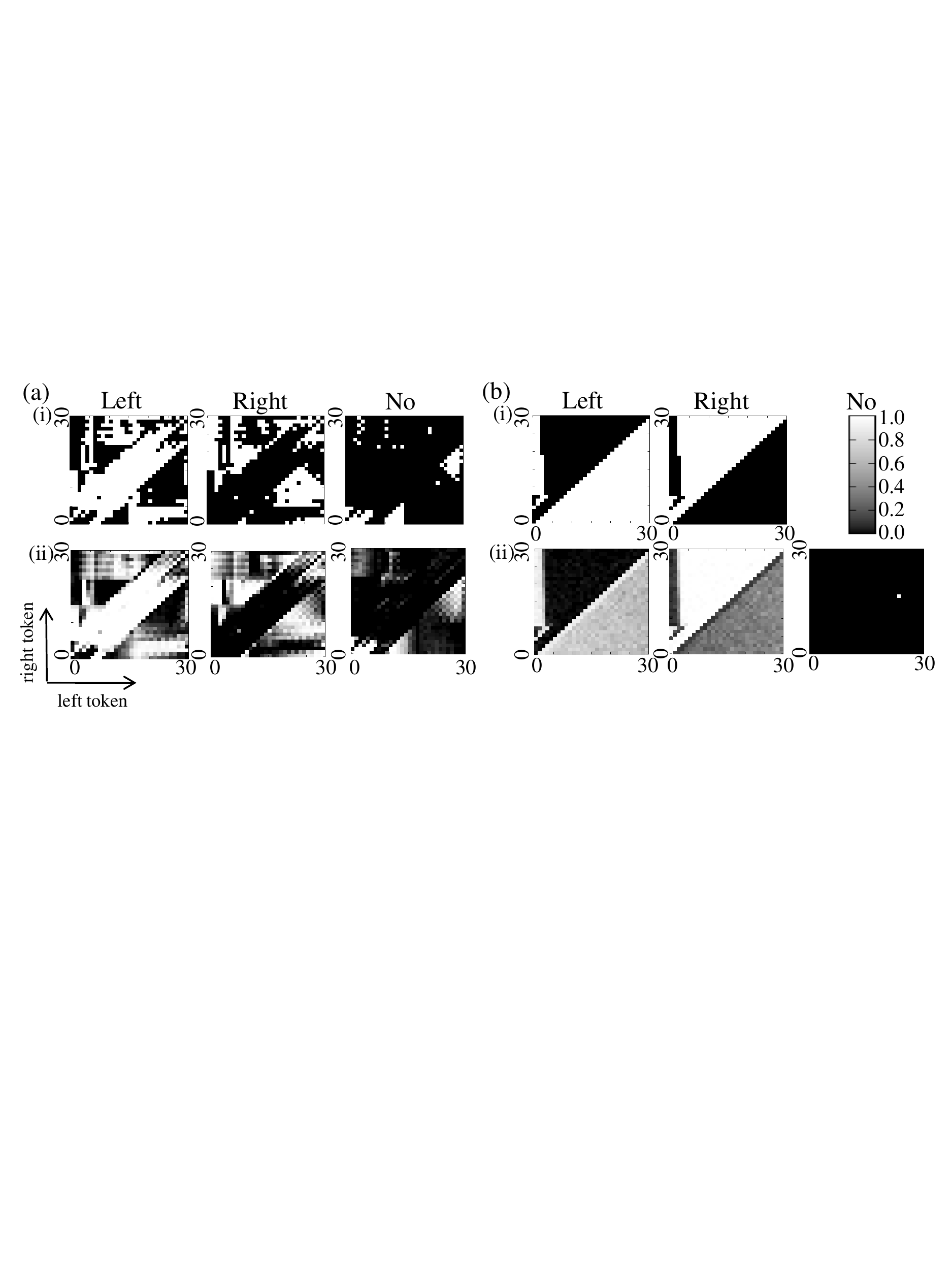}
	\caption{Plots of the probability to get to the end of the left arm, the right arm, and not to get to either of the arms in task 3. For each diagram, the horizontal axis represents the position of the token L, and the vertical axis represents the position of the token R. The left, middle, and right rows represent the probability of getting to the end of the left arm, the right arm, or not getting to either of the arms, respectively. (a) is the result of agent 1, and (b) is that of agent 2. The upper line (i) is the result without noise and the lower line (ii) is the result with noise.}
	\label{fig7}
\end{figure}

In Fig. \ref{fig7}, we saw the relevance of noise to behavior in task 3 for each agent. In Agent 1 (Fig. \ref{fig7} (a)), we can see that it has a tendency to show left judgments when the relative distance between token L and token R is near. For other regions, the judgment to go left, right, or showing no judgment are mixed together. When we apply the noise, the region showing the left judgments, when the relative distance between token L and token R is near, is relatively stable and maintains its judgment, while,  conversely, other regions seem to be influenced strongly. In Agent 2 (Fig. \ref{fig7} (b)), we can see that it has a tendency to show the judgment that corresponds to the token presented earlier to the agent. When we apply the noise, just the judgment that shows left is influenced.

\begin{figure}[htbp]
	\centering
	\includegraphics[width=5.5in, bb=19 297 558 609]{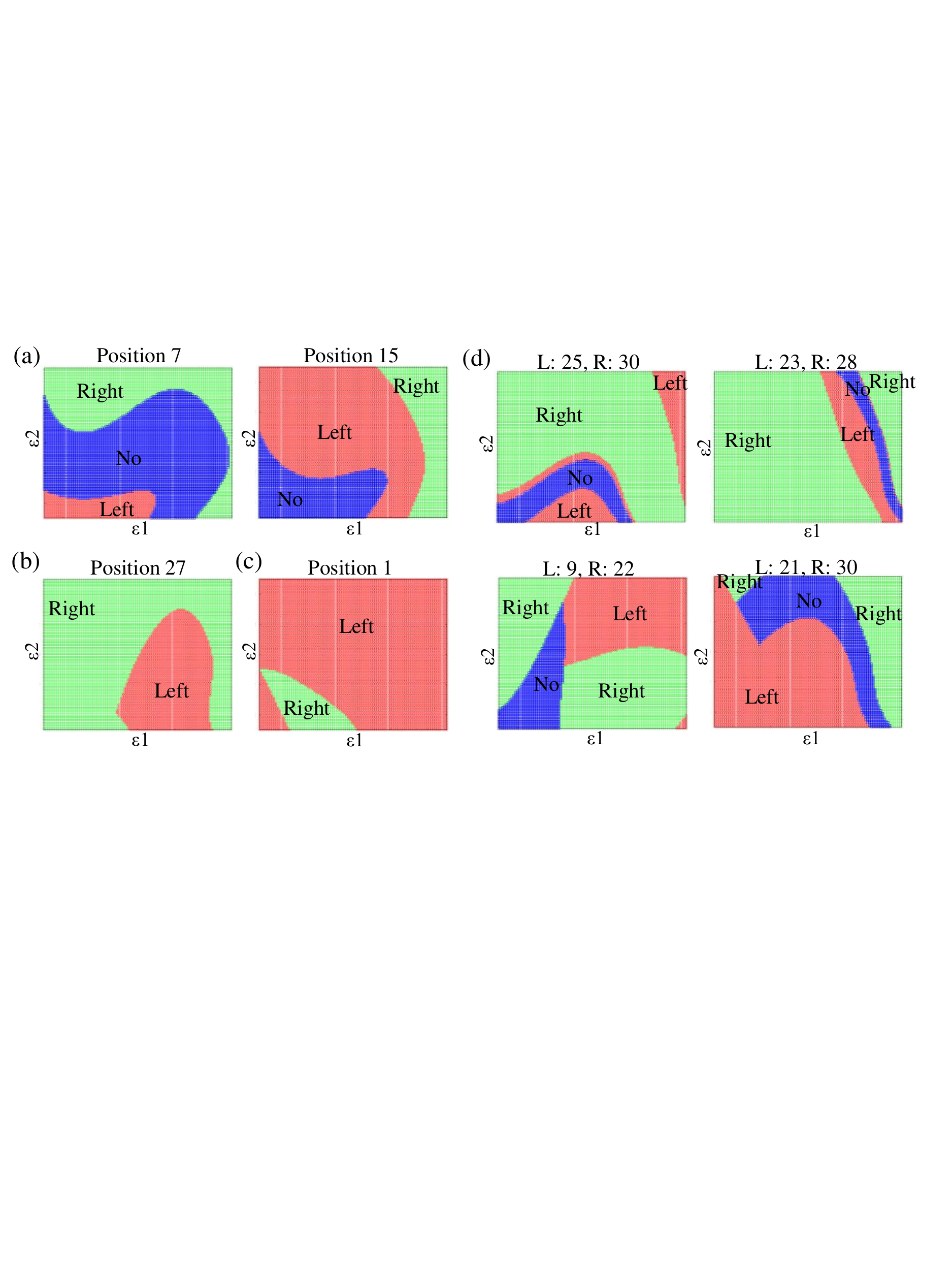}
	\caption{By setting the initial value of two selected recurrent input nodes as $0.5+ \epsilon$, we plotted the decision of the agent. Other recurrent inputs are set to 0.5. For each diagram, the horizontal axis represents the value of $\epsilon$ added to the first recurrent input, and the vertical axis represents the value of $\epsilon$ added to the second one. The range of $\epsilon$ is [-0.5, 0.5]. (a) is the results of Agent 1 for task 1. The left and right diagrams show the case when the token is presented in position 7 and 15, respectively. (b) is the result of Agent 1 for task 2. The diagram shows the case when the token is presented in position 27. (c) is the result of Agent 2 for task 1. (d) is the result of Agent 1 for task 3. The left upper, right upper, left lower, and right lower diagrams show the case where the token L and the token R are presented in position 25 and 30, position 23 and 28,  position 9 and 22, and position 21 and 30, respectively.}
	\label{fig8}
\end{figure}

To see more clearly the relevance of noise in each case, we plotted basins of judgment showing how the different bundles of judgments surround the original trajectory (Fig. \ref{fig8}) \cite{Tani2}. In task 1 and task 2, the basins are clearly separated, but in task 3, we frequently observed a convoluted structure, with one island of judgment in the other island of judgment in the plot. 

\begin{figure}[htbp]
	\centering
	\includegraphics[width=6.5in, bb=23 378 567 632]{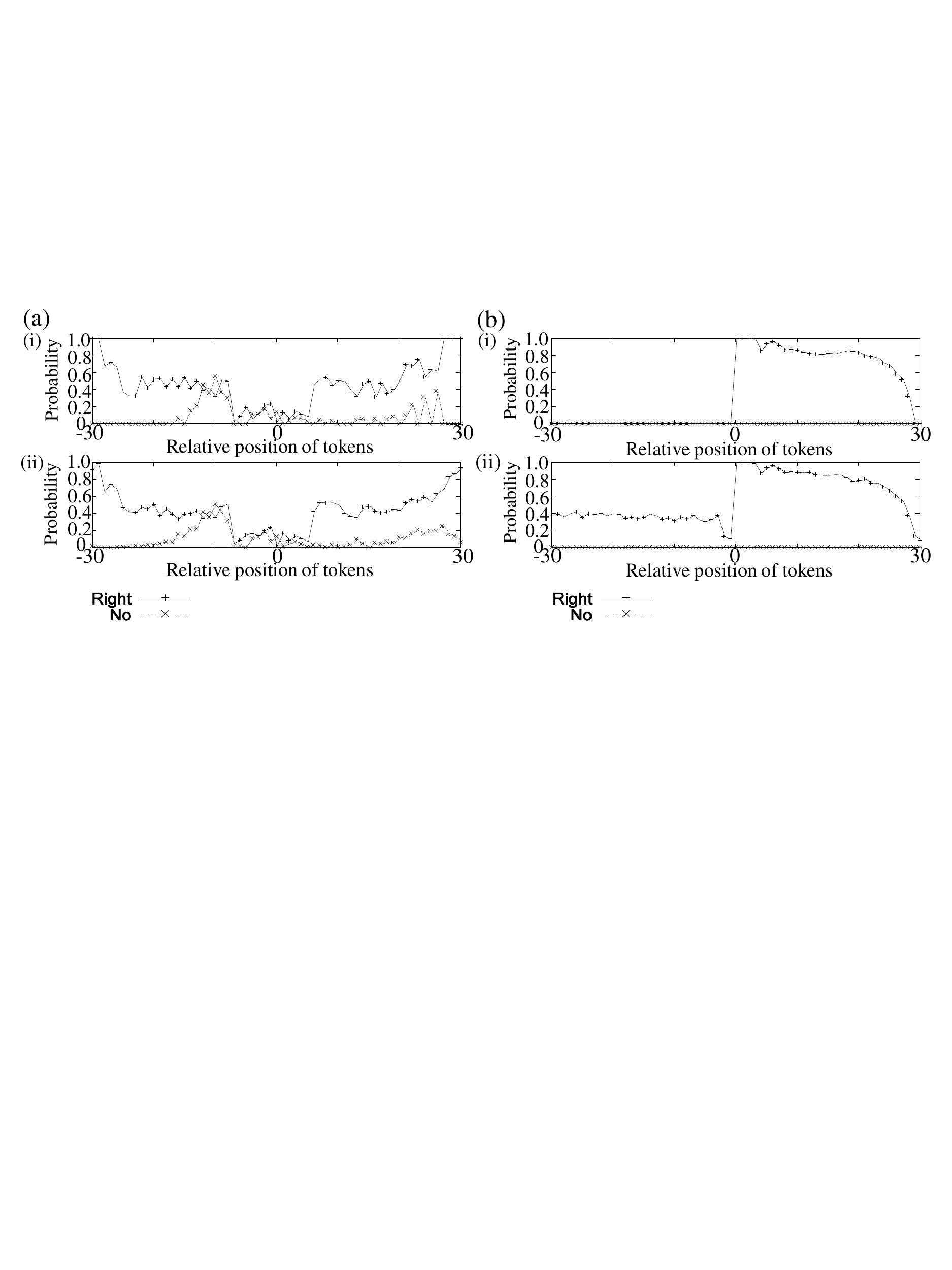}
	\caption{Choice response curves of Agent 1 (a) and Agent 2 (b) for task 3. The upper diagram (i) is the case without noise and the lower (ii) with noise. For each diagram, the horizontal axis represents the relative position of tokens, and the vertical axis represents the probability of choosing the right end of the arm. The relative position of tokens is defined as (the position of token R) - (the position of token L).}
	\label{fig9}
\end{figure}

Fig. \ref{fig9} shows the judgment probability response curves for Agent 1 and Agent 2 in task 3. In Agent 1 (Fig. \ref{fig9} (a)), when the relative distance was from 0 to 8, it goes to the left arm in spite of the ordering of the two tokens. When the relative distance was from 8 to 27, they showed almost half go to the left or right arm. And when the relative distance was from 27 to 30, it goes to the right arm in spite of the ordering of the two tokens. From these observations, we can say that the judgment is based more on the relative distance of the two tokens than on the ordering. And those observed tendencies are less affected by noise, which means its judgment is stable. On the other hand, in Agent 2 (Fig. \ref{fig9} (b)), we can see that it judges its direction to the arm only by the first token it encounters, except when the relative distance of tokens is 30, at which time it goes left. And its judgment is relatively stable, except when the token L was presented first. These are the autonomous choices accompanying our agents.

\section{Discussion}
In this paper, by using the embodied agent, we analyzed its internal dynamics and explored the expression of the agent's autonomous choice. The behavior in which the agents seem to understand the meaning of the symbol is, as a matter of fact, achieved by the sensorimotor loop. The thing we call meaning, or representation in the agent, is not a static and timeless representation, but rather dynamics, which has a certain duration of time, consisting of the interaction between the agent and its environment. In this paper, we focused on this temporal structure and explored the potential implications by analyzing Agent 1 and 2. In Agent 1, since it was not fully evolved, we observed that according to the type of token, when the presented token was perceived, and its internal dynamics, sometimes the performance fluctuated. In this case, since the performance of the agent depends on the specific situation, we can say that its generalization is shallow and we cannot assume a static representation, which will not change its property despite the change in the situation. On the other hand, in Agent 2, since it was evolved enough, its internal dynamics were relatively stable, so that it showed almost the same pattern of response whenever it perceived the tokens. Also, its performance to the presented tokens was almost perfect. In this case, it might be possible to assume a static representation and say that it generalized the meaning of tokens well. Next, by using the same agents, we observed what happened if two types of tokens were presented. As a result, we showed that Agent 1 was using the relative distance between the tokens, rather than a spatial order, to achieve an autonomous choice. On the other hand, Agent 2 was specifically sensitive to the first token it perceived. These results are difficult to speculate on when regarding the performance of tasks that only present a single token. Thus, we can suggest that if we assume that what we call meaning is sustained in a specific duration of time, it is important to investigate the interaction between the durations of time that characterizes the meaning for each. Moreover, this interaction would not be like an operation defined between symbols. In this paper, we presented one example showing the importance of this aspect. This should be further explored in future work.

Recently, in neuroscience, similar kinds of issues are presented in different contexts. They are related to various problems, such as how time perception is achieved \cite{clock} and its relation with its body image \cite{Kitazawa,Kohei1} in the brain. Recent findings suggest that it is not realized as a static clock, but rather a dynamic one that changes its properties constantly according to external stimuli \cite{clock}. Since biological systems do not have an external tuner for their clocks, but should adjust them by themselves, if we consider how they self-organize and self-tune their clocks, we should take into account the interaction between the system and their environment through their body. This would also be an interesting direction for exploration in future work. 
\bibliographystyle{aipproc}   

\bibliography{Kohei}

\IfFileExists{\jobname.bbl}{}
 {\typeout{}
  \typeout{******************************************}
  \typeout{** Please run "bibtex \jobname" to optain}
  \typeout{** the bibliography and then re-run LaTeX}
  \typeout{** twice to fix the references!}
  \typeout{******************************************}
  \typeout{}
 }
\clearpage
\clearpage
\clearpage
\clearpage
\clearpage
\clearpage
\clearpage
\clearpage
\clearpage
\end{document}